\documentclass[revtex4]{emulateapj}

\usepackage{longtable,lscape}
\usepackage{natbib}
\usepackage{graphicx}
\usepackage{float}

\usepackage[caption = false]{subfig}
\usepackage{changepage}
\usepackage{amsmath} % or simply amstext

\def\m2s2{\,m$^{2}$\,s$^{-2}$} %m2.s -2
\begin{document}
\newcommand{\motz}[1]{{\sffamily\bfseries #1}}
\newcommand{\sub}[1]{\ensuremath{_\mathrm{#1}}}

\shorttitle{New sub-stellar companion around HD 284149}
\shortauthors{M. Bonavita et al.}

   \title{A new sub-stellar companion around the young star HD 284149}

   \author{Mariangela Bonavita \altaffilmark{1}\email{mariangela.bonavita@oapd.inaf.it},
   Sebastian Daemgen \altaffilmark{2}, Silvano Desidera \altaffilmark{1}, Ray Jayawardhana \altaffilmark{2,3}, Markus Janson \altaffilmark{3}, David Lafreni{\`e}re \altaffilmark{4}} 

\altaffiltext{1}{INAF -- Osservatorio Astronomico di Padova, Vicolo dell'Osservatorio 5, I-35122, Padova, Italy}
\altaffiltext{2}{Department of Astronomy and Astrophysics, University of Toronto, Toronto, ON, Canada}
\altaffiltext{3}{Department of Physics and Astronomy, York University, Toronto, ON L3T 3R1, Canada}
\altaffiltext{4}{Astrophysics Research Centre, Queen's University Belfast, Belfast, UK }
\altaffiltext{5}{Department of Physics, University of Montreal, Montreal, QC, Canada}

\begin{abstract}
Even though only a handful of sub-stellar companions have been found via direct imaging, each of these discoveries has had a tremendous impact on our understanding of the star formation process and the physics of cool atmospheres. Young stars are prime targets for direct imaging searches for planets and brown dwarfs, due to the favorable brightness contrast expected at such ages and also because it is often possible to derive relatively good age estimates for these primaries. Here we present the direct imaging discovery of HD 284149 b, a $18-50~M_{Jup}$ companion at a projected separation of ∼ 400 AU from a young ($25^{+25}_{−10}$~Myr) F8 star, with which it shares common proper motion
\end{abstract} 


\section{Introduction}
In recent years, the rapid improvement of high-contrast imaging instrumentation and techniques have led to the 
discovery of a number of wide sub-stellar companions to nearby young stars, down to planetary mass 
\citep[e.g.][]{2005A&A...438L..29C, 2006ApJ...649..894L,2008Sci...322.1348M, 2010Natur.468.1080M,2009A&A...493L..21L, 2014ApJ...780L..30C}. Several of these discoveries, such as AB Pic~b \citep{2005A&A...438L..29C}, HN~Peg~B \citep{2007ApJ...654..570L}, 1RXS J1609~b \citep{2010ApJ...719..497L, 2008ApJ...689L.153L},   HIP 78530~b \citep{lafreniere2011}  and the recently discovered HD~106906~b \citep{2014ApJ...780L...4B} and  ROXS~42B~b \citep{2014ApJ...780L..30C},
have mass ratios with respect to their parent stars of only $\sim 1\%$ and 
seriously challenge the current planet formation paradigm.
In particular, their large separations are hard to explain and suggest they might be extreme outcomes of their underlying formation mechanism, regardless of whether it is based on core accretion or disk instability.

Our previous survey of 91 stars in the USco region \citep{2014ApJ...785...47L} implies a frequency of wide companion for such regions of 4-5\%, in agreement with other studies \citep{2011ApJ...726..113I}.
This suggests a frequency of wide companions in star forming regions comparable to the values for young moving groups or the field, reported for example by \cite{2007ApJ...670.1367L, 2009ApJS..181...62M, 2010A&A...509A..52C}.

Most recently we also confirmed three new companions with masses of $\sim 40-100~M_{Jup}$ and separations of 
$\sim 40-230~AU$ in the Scorpius-Centaurus (Sco-Cen) region \citep{2012ApJ...758L...2J}.
These companions represent an interesting intermediate between stellar companions and the $\sim 10-20~M_{Jup}$ ones described above in the Upper Scorpius (USco) region. 
The existence of such a seemingly continuous population might imply that binary formation extends all the way down to planetary
masses for wide separations, or at least that mass alone is not a clear-cut diagnostic for distinguishing 
between formation mechanisms. 
In order to further address these issues, we conducted a survey of 74 stars in the Taurus star forming region with ALTAIR/NIRI \citep{2000SPIE.4007..115H,2003PASP..115.1388H}.
The results of the full survey will be presented in a dedicated paper (Daemgen et al. 2014, ApJ Submitted).
Here we present the discovery of a $18-50~M_{Jup}$ companion at a projected separation of $\sim 400$~AU from the F8 star HD~284149. 
A dedicated analysis of the host properties is also presented in Sec.~\ref{sec:host}, addressing the question of its questionable Taurus membership.

\section{Observations and data reduction}
\label{sec:obsred}
HD\,284149 was observed during six epochs between October 2011 and March 2014 on Gemini North with the adaptive-optics assisted NIRI instrument (Hodapp et al. 2003) in J, H, and Ks-band. The f/32 camera provided a sampling of 21.9~mas/pixel and a field of view (FoV) of 22\arcsec$\times$22\arcsec. Total integration times varied between 9\,sec (J) and $\sim$7\,sec (Ks) and were taken as a series of coadds in a 5-point dither pattern to increase dynamic range and allow sky subtraction.
The details of the observing times for each epoch, together with the mean airmass and seeing at each observing date are reported in Tab.~\ref{tab:comp_char}.
After subtraction of a striping pattern frequently observed in NIRI images, all images were flat fielded, bad pixel corrected, and sky subtracted. The field distortion was corrected as described in \cite{2014ApJ...785...47L} who determine a residual astrometric uncertainty of 15\,mas, 25\,mas, and 50\,mas at radii 4\arcsec, ~8\arcsec, ~and ~12\arcsec\ ~from the center, respectively.

The left panel of Fig.~\ref{fig:cpm} shows one of the fully reduced images of HD\,284149 and its companion obtained with NIRI in 2012B. The achieved full width at half maximum of the point spread function is 0\farcs08, and the companion, at a separation of $3.7\arcsec$ is detected at  $\gtrsim$14$\sigma$.
As part of our survey for faint companions in Taurus (Daemgen et al. 2014, subm.), we also obtained deep exposures of HD\,284149 in H and J band, which confirm the presence of the companion with high S/N $>$ 200. These observations, however, saturate the central star and render the relative astrometry and photometry less precise than in the lower-S/N images analyzed here, and are not further used.

\begin{figure*}[]
\centering
\includegraphics[width=110mm]{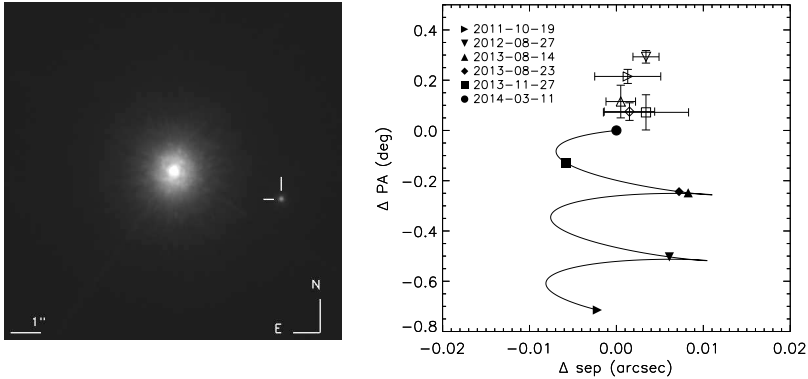}\label{fig:cpm}
\caption{ {\bf Left:} Candidate companion detected in $K_s$ with GEMINI/NIRI in one of the images obtained in 2012B. {\bf Right:} Evaluation of common proper motion for the detected companion to HD 284149. The continuous line shows the motion of a background star and the filled symbols its positions of at the various epochs. The open symbols represent the corresponding measurements of the position of HD 284149 B at the same epochs. If the star and companion are co-moving, the open symbols with error bars should be close to the filled circle in the origin, which represents our most recent epoch used as reference. If the companion candidate does not move with respect to the background, then the open symbol is consistent with the location of the identical filled symbol.}
\label{fig:cpm}
\end{figure*}

\section{Host star properties}%Re-evaluation of the system age
\label{sec:host}
HD 284149 was included among the members of Taurus-Auriga association by \cite{1996A&A...312..439W}, but it is not considered in the compilations by \cite{2008hsf1.book..405K} and \cite{2014ApJ...784..126E}. Therefore, a re-assessment of the stellar properties is needed. HD 284149 was classified as F8 by \cite{2012ApJ...745..119N} and G1 by \cite{2000A&A...359..181W}.

A young age of the star is supported by the large lithium EW \citep{2000A&A...359..181W}, the large X-ray luminosity as revealed by ROSAT, the photometric variability and fast rotation \citep[a period of 1.079 days is reported by ][]{2007IBVS.5752....1G}. The short-term RV monitoring by \cite{2012ApJ...745..119N} was able to exclude the possibility of a tidally-locked binary. However with a 3 Km/s difference between mean RV from \cite{2000A&A...359..181W} and \cite{2012ApJ...745..119N} being about 3 km/s, a binarity with periods of months or years cannot be excluded\footnote{In order to reduce the impact of possible binarity on the space velocity, we decided to use the average velocity. 
Efforts are underway to better understand the multiplicity status of HD 284149 and will be presented in further publications. There are no indication from non-detection on line profile alteration in Nguyen et al. 2012 and lack of direct detection in our own images  that the companion contributes significantly to the integrated flux. The expected impact on the age indicators is then minor.}

\begin{figure*}[]
\centering
\includegraphics[width=110mm]{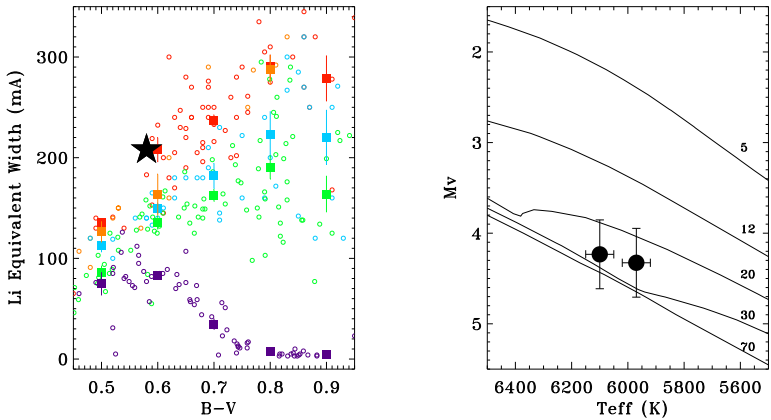}
\caption{{\bf Left:} Lithium equivalent width vs B-V of HD284149 compared to that of members of nearby groups and clusters of young age. The black star represents our target, orange circles represent members of beta Pic moving gropus and TW Hya association (age 8-16 Myr) red circles members of Tucana, Columba and Carina MG (age 30 Myr);
light blue circles: members AB Dor MG (age 100 Myr); green circles: Pleiades OC (age 125 Myr); purple circles: Hyades OC (age 625 Myr). The filled squares represent the median values of EW Li for the corresponding color bin.
 Li EW data are from \cite{2006A&A...460..695T} for young assocition, \cite{1993AJ....106.1059S} for the Pleiades, and \cite{1993ApJ...415..150T} for the Hyades.
{\bf Right:} Position of HD284149 (filled dots) on HR diagram for the temperatures corresponding to F8 and G1 spectral classification. Overplotted are the 5, 12, 20, 30, and 70 Myr isochrones from \cite{2012MNRAS.427..127B}.}
\label{fig:age}
\end{figure*}

From the G1-F8 spectral classification, an effective temperature of 5970-6100 K is derived following \cite{2013ApJS..208....9P}. Photometric colors are broadly consistent with such temperatures, with a detailed comparison hampered by obervational scatter (e.g. peak-to valley differences larger than 0.2 mag in V band), possibly linked to the photometric variability of the star. Adopting the V magnitude from ASAS ($9.653\pm0.060$), the $V-K_s$ color is 1.55 mag. Comparison with the pre-main sequence (pre-MS) intrinsic colors of young stars by \cite{2013ApJS..208....9P} suggests a reddening E(B-V) of about 0.05-0.08 mag for a G1 and F8 star, respectively.
Slightly smaller amounts of reddening are indicated by the B-V and V-I colors. Such amount of reddening is not unexpected at the distance of the star. The trigonometric parallax from \cite{2007A&A...474..653V} is $9.24\pm1.58$ mas.  A comparison with members of young moving groups (MGs) (see left panel of Fig.~\ref{fig:age}) indicates that the lithium equivalent width of HD 284149 \citep[208 m\AA,][]{2000A&A...359..181W} is comparable with that of members of $\beta$ Pic, Tuc-Hor, Columba and Carina moving groups of similar temperatures, and clearly above that of Pleiades open cluster and AB Dor moving group.
A similar result is obtained for a comparison of the X-ray luminosity ($\log L_{X}/L_{bol}=-3.3$ for HD 284149). 

The position of HD 284149 on an HR diagram (see right panel of Fig.~\ref{fig:age}) is close to the 25 Myr isochrone using the theoretical models by \cite{2012MNRAS.427..127B}, with ages between 15 to 100 Myr also compatible with the data. %A stellar mass of $1.14\pm0.05~M_{\odot}$ is derived. 
This isochronal age is on average older than that obtained by other authors \citep[14-16 Myr, ][]{2000A&A...359..181W,palla2002} because of the revision in the trigonometric parallax, with respect to the value reported in the Hipparcos catalogue \citep{1997ESASP1200.....P}. 

The resulting kinematic parameters, adopting the \cite{2007A&A...474..653V} parallax and proper motion and the mean of the RVs obtained by \cite{2000A&A...359..181W} and \cite{2012ApJ...745..119N}, are $U=-12.3$~km/s, $W=-6.4$~km/s and $W=-8.8$~km/s. This is similar to that of the Octans association discussed in \cite{2008hsf2.book..757T}, whose proposed members are however all very far from HD 284149 on the sky.
Recently \cite{2013ApJ...778....5Z} identified an additional group of young stars with similar kinematics to the Octans association but with a smaller distance from the Sun and a different sky distribution, labeled as Octans-Near. While the link between Octans and Octans-Near groups and the existence of the latter as a true moving group deserves further investigation, we note that one of the Octans-Near proposed members, HIP 19496, is separated on the sky by about 5 deg from HD 284189, it has a comparable distance (98 vs 108 pc), and the space velocities of the two stars differ by just 2.7 km/s. \cite{2013ApJ...778....5Z} estimate an age of 30 Myr for HIP 19496, similar to our determination for HD 284189.

In summary, HD 284149 is significantly older than the bulk of the Taurus-Auriga association. Membership is still possible in case of earlier start of star formation in the outer regions of the association \citep[][]{2002ApJ...581.1194P}. Independently of the Tau-Aur membership, the Li EW well above the Pleiades locus coupled with the position on HR diagram above ZAMS and other age diagnostics indicate an age of about 25~Myr, with minimum and maximum values of about 15 and 50~Myr respectively.
A summary of the stellar parameters is given in Table~\ref{tab:properties}.

\begin{table}[ht!]
\small
\begin{tabular}{l|l|l|l}
\hline
\hline
Parameter           &  Host Star                    & Companion                     & Ref.  \\
\hline
\hline
V (mag)             &  9.653$\pm$0.060              & & 2     \\
B-V                 &  0.58                         & & 3     \\
V-I                 &  0.675$\pm$0.088              & & 4     \\
J (mag)             &  8.479$\pm$0.043              & 15.516$\pm$0.043            & 5,1     \\
H (mag)             &  8.208$\pm$0.021              & 14.715$\pm$0.047            & 5,1     \\
K (mag)             &  8.100$\pm$0.029              & 14.332$\pm$0.040            & 5,1     \\
parallax (mas)      &  9.24$\pm$1.58                & & 6     \\
E(B-V)              &  0.05$\pm$0.05                & & 1     \\
RV (Km/s)           &  14.0$\pm$2.0                 & & 7,8   \\
Ew Li m($\AA$)      &  208                          & & 7     \\
Prot (d)            &  1.079                        & & 9     \\
log Lx/Lbol         &  -3.3$\pm 0.1$               & & 1     \\
vsini (Km/s)        &  27.0$\pm$1.9                 & & 8     \\
Age (Myr)           &  $25^{+25}_{-10}$             & & 1     \\
Sp. Type            &  G1-F8                        & M8-L1                         & 7,8,1  \\
$T_{Eff}$~(K)       &  5970-6100                    & $2537^{+95}_{-182}$         & 1     \\
Mass                & $1.14\pm0.05~M_{\odot}$       & $32^{+18}_{-14}~M_{Jup}$  & 1     \\

\hline
\hline
\end{tabular}
\caption{ Summary of properties of both HD 284149 and its companion.\\
{\bf References:} [1] This paper;  [2] ASAS \citep{2002AcA....52..397P}; [3] SIMBAD; [4] TASS \citep{2000PASP..112..397R}; [5] 2MASS \cite{2003yCat.2246....0C}; [6] \cite{2007A&A...474..653V};  [7] \cite{2000A&A...359..181W}; [8] \cite{2012ApJ...745..119N}; [9] \cite{2007IBVS.5752....1G} }
\label{tab:properties}
\end{table}

\section{Companion properties}

The relative position of HD~284149 and its companion were determined with PSF photometry using \emph{daophot} in \emph{IRAF}. The bright star HD~284149 was used as PSF reference to obtain relative photometry and astrometry of the companion. Statistical uncertainties were inferred from the rms noise between the individual dither exposures for each epoch and filter. Systematic flux uncertainties are estimated from the residuals after PSF subtraction to be $\lesssim$5\%, and systematic astrometric uncertainties are dominated by the uncertainty of the distortion correction at the position of HD~284149b of $\lesssim$15\,mas. The resulting astrometry and photometry are listed in Tab.~\ref{tab:comp_char}, while a summary of the derived properties is given in Tab.~\ref{tab:properties}. 

The right panel of Fig.~\ref{fig:cpm} shows the relative change of separation and position angle of the companion between our previous observations (filled right-facing triangle) with respect to the most recent one (filled circle). 
We conclude that the point source we imaged is consistent with a co-moving companion at $\sim 400$~AU from HD284149 with $>99\%$ confidence according to a $\chi^2$ test.

As discussed in Sec.~\ref{sec:host}, the age of this system is controversial, as it appears to be older than other members of the Taurus association.
With an adopted age of  $25^{+25}_{-10}~Myr$, the Ks brightness of the companion suggests a mass of $32^{+18}_{-14}~M_{Jup}$, according to the DUSTY models by \cite{2000ApJ...542..464C}. 
Using the DUSTY models by \cite{2000ApJ...542..464C} we derived an effective temperature of $2337^{+95}_{-182}$~K. Together with the color measurements (J-H~$\sim 0.8$, H-K~$\sim 0.4$) this suggests a spectral type between M8 and L1 \citep[see e.g.][and www.pas.rochester.edu/$\sim$emamajek/ for the extended table]{2013ApJS..208....9P}, but further measurements are required to better constrain this.

\begin{figure*}[]
\centering
\includegraphics[width=110mm]{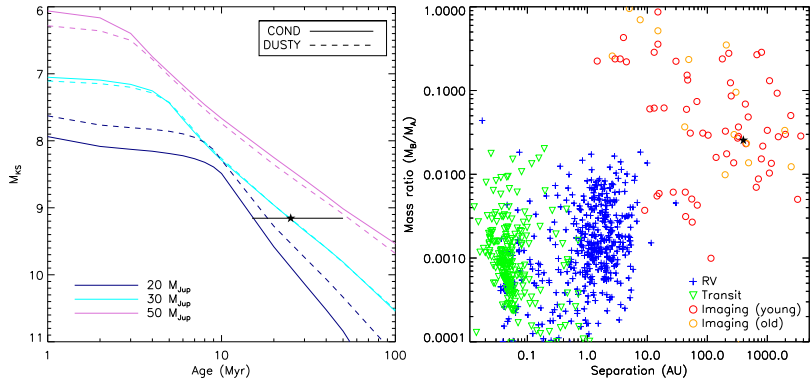}
\caption{ {\bf Left:} Absolute magnitude in $K_S$ band vs age of a companion of 20, 30 and 50~$M_{Jup}$ (blue, cyan and purple curve, respectively) according to the COND (solid lines) models by \cite{2003A&A...402..701B} and DUSTY (dashed lines) models by \cite{2000ApJ...542..464C}. The position of HD 284149 B is marked by a filled star. {\bf Right:} Mass ratio vs separation of HD 284149 (filled star), compared to those of other known low-mass companions discovered so far. The companions discovered using the radial velocity (RV) technique are marked with blue crosses, the ones transiting their parent stars with green triangles. Finally, directly imaged companions are represented by red circles if the stellar age is less than 500~Myr (young companions), and with red circles otherwise.} 
\label{fig:age_mag}
\end{figure*}

\begin{table*}[]
\tiny
\centering
\begin{tabular}{l|ll|lll|lll|ll|lll}
\hline \hline
UT Date     & \multicolumn{8}{|c|}{Observing conditions and setup}                                              & \multicolumn{2}{|c|}{Astrometry} & \multicolumn{3}{|c}{Photometry} \\ \cline{2-14}
            & Seeing & Air Mass & \multicolumn{3}{|l|}{Total Exposure Time} &  \multicolumn{3}{|l|}{$N_{Coadds}$} &  Sep.      & PA     & $\Delta~J$ & $\Delta~H $       & $\Delta~K_S$ \\
            & (arsec)&      & J (sec)  & H (sec)  & K$_S$ (sec) &  J   & H  & K$_S$             &  (arcsec)  & (deg)  & (mag)      & (mag)             & (mag) \\
\hline
\hline
2011-10-19 & 1.086  & 0.557 & --   & --   & 10.50 & -- & -- & 30 & 3.6826 $\pm$ 0.0038 & 255.065 $\pm$ 0.028 &   --               &   --              & 6.208 $\pm$ 0.031 \\
2012-08-27 & 1.018  & 0.417 & --   & --   & 8.00  & -- & -- & 40 & 3.6847 $\pm$ 0.0015 & 255.143 $\pm$ 0.025 &   --               &    --             & 6.171 $\pm$ 0.041 \\
2013-08-14 & 1.241  & 1.061 & --   & --   & 7.00  & -- & -- & 20 & 3.6818 $\pm$ 0.0017 & 254.965 $\pm$ 0.065 &   --               &    --             & 6.319 $\pm$ 0.045 \\
2013-08-23 & 1.053  & 0.378 & 9.10 & --   & 7.00  & 26 & -- & 25 & 3.6828 $\pm$ 0.0029 & 254.925 $\pm$ 0.032 & 6.99  $\pm$ 0.080  &    --             & 6.218 $\pm$ 0.085 \\
2013-11-27 & 1.036  & 0.658 & 8.96 & 8.32 & 7.00  & 28 & 26 & 35 & 3.6847 $\pm$ 0.0049 & 254.922 $\pm$ 0.070 & 6.990 $\pm$ 0.047  & 6.647 $\pm$ 0.061 & 6.262 $\pm$ 0.058 \\
2014-03-11 & 1.267  & 0.716 & 9.00 & 8.32 & 7.00  & 10 & 26 & 35 & 3.6813 $\pm$ 0.0038 & 254.850 $\pm$ 0.157 & 7.06  $\pm$ 0.280  & 6.738 $\pm$ 0.059 & 6.299 $\pm$ 0.074 \\
\hline
\hline
\end{tabular}
 \caption{Detailes of the observations setup and conditions, and relative astrometry and photometry of HD 284149 and its companion}
 \label{tab:comp_char}
\end{table*}

\section{Discussion and Conclusions}

We presented here the detection of a substellar (~32~$M_{Jup}$ assuming an age of 25~Myr, see Fig.~\ref{fig:age_mag}~a and Tab.~\ref{tab:properties}) companion orbiting the young star HD 284149 at a separation of $\sim 400$~AU. 

Fig.~\ref{fig:age_mag} shows a comparison of HD 284149 mass-ratio and separation with the values of the planetary companions found by radial velocities (RV) and transit methods, as well as other directly imaged planetary and brown dwarf companions. 
The group of objects with separation $<100~AU$ and mass ratio $<0.01$ seems to be well separated from the one including companions with larger separation and mass ratio, suggesting that different formation mechanisms could be at play. 
The small mass ratio of the first group might suggest planet-like formation, but objects with similar mass-ratios at larger separations are difficult to explain. 

HD 284149b shows very similar properties to objects like ROXs 42Bb or AB Pic b which, as suggested by \cite{2014ApJ...780L..30C}, places it between the bona-fide planets and the lowest mass brown dwarfs imaged so far. 
Together with these and other companions of similar mass and separation, such as HN~Peg~B and HD~106906,  HD~284149b represents a challenge for our understanding of the formation of low-mass companions at very wide separations.
The high mass ratio of these systems might suggest a planet-like origin, but at the same time, their estimated mass is well above the deuterium burning limit, suggesting a stellar-like formation. 
 
The existence of such companions suggests that mass ratio alone is not sufficient to distinguish between planet-like and star-like formation, at least for wide companions \citep[see also][]{2012ApJ...745....4J}. 

Finally, the findings of dedicated RV campains around young stars seem to suggest a paucity of close-in planetary companions around these targets. Only few close in companions have been detected around young early-G and F-type stars, such as HD 70573 \citep{2007ApJ...660L.145S} and HD 113337 \citep{2014A&A...561A..65B}. 
Their small number seem to imply a lower frequency of such companions if compared to the more massive, more distant ones as HD 284149B.
If confirmed, this could suggest that multiple planet formation mechanisms are at play around these objects. 

The object brightness and separation makes HD~284149~b a very well suited target for detailed characterization of both the host star and the companion. Efforts toward this direction are already under way and will be presented in further publications.

\subsubsection*{Acknowledgments}

{ This work was supported by grants from the NSERC of Canada and the University of Toronto McLean Award to R.J. SD is supported by a McLean Postdoctoral Fellowship. M.B. is founded through the “Progetti Premiali” funding scheme of the Italian Ministry of Education, University, and Research, and through the PRIN-INAF 2010 “Planetary systems at young ages and the interactions with their active host stars”. Based on observations obtained at the Gemini Observatory, operated by the Association of Universities for Research in Astronomy, Inc., under a cooperative agreement with the NSF on behalf of the Gemini partnership. This research has made use of the SIMBAD database, operated at CDS, Strasbourg, France}

%%\bibliographystyle{aa}
%%\bibliography{ddt}

\end{document}